\documentclass[journal, 12pt]{IEEEtran}
\onecolumn
\usepackage{setspace}

\usepackage{amsmath}  
\usepackage{amsfonts}  
\usepackage{amssymb}   
\usepackage{caption}
\usepackage{graphicx}  

\usepackage[T1]{fontenc}  
\usepackage{lmodern}

\usepackage{hyperref}  

\usepackage[utf8]{inputenc}  

\usepackage{array}

\usepackage{listings}

\usepackage{enumitem}

\usepackage{multicol}

\usepackage[margin=1in]{geometry}

\usepackage{mathtools}

\usepackage{float}  
\usepackage{algorithm}
\usepackage{algpseudocode}
\usepackage{graphicx}
\ifCLASSINFOpdf
\else
\fi
\begin{document}
%

\title{2D-DOA Estimation and Auto-Calibration in UCAs via an Integrated Wideband Dictionary}
%
%
%

\author{Zavareh Bozorgasl\thanks{Z. Bozorgasl .}, Hao~Chen,~\IEEEmembership{Member,~IEEE}  \thanks{H. Chen is with the Department of Electrical and Computer Engineering, Boise State University, Boise,
ID, 83712 (email: haochen@boisestate.edu) }
        Mohammad J. Dehghani\thanks{M. J. Dehghani is with the Department of Electrical and Electronics Engineering, Shiraz University of Technology, Shiraz, Iran (email: dehghani@sutech.ac.ir).}
}

\maketitle

\begin{abstract}
In this paper, we present a novel auto-calibration scheme for the joint estimation of the two-dimensional (2-D) direction-of-arrival (DOA) and the mutual coupling matrix (MCM) for a signal measured using uniform circular arrays. The method employs an integrated wideband dictionary to mitigate the detrimental effects of the discretization of the continuous parameter space over the considered azimuth and elevation angles. This leads to a reduction of the computational complexity and obtaining of more accurate DOA estimates. Given the more reliable DOA estimates, the method also allows for the estimation of more accurate mutual coupling coefficients. The method utilizes an integrated dictionary in order to iteratively refine the active parameter space, thereby reducing the required computational complexity without reducing the overall performance. The complexity is further reduced by employing only the dominant subspace of the measured signal. Furthermore, the proposed method does not require a constraint on the prior knowledge of the number of nonzero coupling coefficients nor suffer from ambiguity problems. Moreover, a simple formulation for 2-D non-numerical integration is presented. Simulation results show the effectiveness of the proposed method.
\end{abstract}

\textbf{Index Terms---}Direction-of-Arrival, Sensor Array Processing, Sparse Representation, Integrated Dictionary, Auto-Calibration, Uniform Circular Array, Mutual Coupling Matrix.

%
\IEEEpeerreviewmaketitle

\section{Introduction}
%
%
%
%

Estimating the direction-of-arrival (DOA) of a plane wave impinging on a sensor array, a classical problem in array signal processing, has been widely investigated in recent decades \cite{ref1, ref2, ref3}. Although the literature mostly deals with DOA estimation techniques which use the assumption of a perfectly known measurement array, the resulting steering vector will generally only be approximately known. Indeed, the manifold will suffer from effects such as mutual coupling, which may cause a substantial degradation of the performance of applied algorithms \cite{ref4, ref5, ref6}. Two of the most commonly used array configurations are the uniform linear array (ULA) and the uniform circular array (UCA). As the mutual coupling effect will be more pronounced for a circular than for a linear array, the detrimental effects of such coupling will affect the resulting performance more, suggesting the current study.

In order to alleviate the mutual coupling effects, two main categories of algorithms have been proposed in the literature. The first of these employ offline calibration algorithms\cite{ref8, ref9}, which require calibration sources, i.e., some sources with perfectly known locations, to determine the array manifold matrix. In contrast, the second category uses auto-calibration or online calibration algorithms\cite{ref10, ref11, ref12,ref13}, which do not require such source calibration, instead aiming to compensate for the calibration automatically. As finding additional calibration sources is not possible in many practical applications, the auto-calibration algorithms are often preferable.

Moreover, as the discretization of the parameter space leads to undesirable off-grid effects, there has recently been increasing interest in continuous parameter space techniques\cite{ref15, ref16}. One such alternative is to use a continuous dictionary together with an atomic norm penalty\cite{ref17, ref18, ref19}. While such an approach often yields an accurate signal reconstruction, this form of problem formulation often leads to optimization problems with a high computational complexity. In this work, instead, we propose the use of an integrated wideband dictionary, as introduced \cite{ref20}, which is formed over subsets of the continuous parameter space. The essence of using this kind of dictionary is the ability to discard the non-activated subsets and preserving, then refining the activated ones for further zooming without the risk of missing components. These steps are then iterated until one achieves the desired resolution.

Common problems of earlier auto-calibration, including ambiguity problems \cite{ref12, ref13} and/or the requirement of a priori knowledge of the number of nonzero mutual coupling coefficients \cite{ref21, ref22}, both typically being infeasible in practice. The here proposed method, which iteratively estimates both the DOAs and the mutual coupling coefficients, suffers from neither of these shortcomings. For the DOA estimation stage, instead of using a conventional grid-based dictionary which would necessitate a high computational complexity to allow for a sufficiently fine grid, a wideband dictionary framework is employed. By forming the used dictionary consisting of subsets of the continuous parameter space, and applying the noted screening process, the non-activated sets may be discarded in the first stage, whereas the activated subsets are retained and refined for further processing. In this sense, the initial DOA estimation stage will constitute an initial coarse estimate that will not suffer from the usual off-grid effects, such as missing active components. Via the further steps, i.e., refining the remaining activated bands, the closely spaced DOAs are separated. Next, the MCM estimation is formed utilizing the complex symmetric circular Toeplitz structure of the MCM, as was proposed for 1-D DOA estimation in \cite{ref23, ref24}. Here, this mutual coefficient estimation stage is further extended to the 2-D DOA estimation problem for the UCA configuration.

The remainder of the paper is organized as follows. Section \ref{sec:data model} outlines the data model for 2-D UCAs. Section \ref{sec:proposed method} describes the details of the new method for joint iterative 2-D DOA and MCM estimation. Section \ref{sec:simulations} provides some simulations to demonstrate the effectiveness of the proposed method. Finally, Section \ref{sec:conclusion} concludes the work.

\textbf{Notation:} Bold lower-case (upper-case) letters are used throughout to denote vectors (matrices). $\mathbb{C}$ denotes the set of complex numbers. $(\cdot)^T$ and $(\cdot)^H$ represent the transpose and the complex conjugate transpose, respectively. The notations $\|\cdot\|_k$ and $\|\cdot\|_F$ stand for the k-norm of a vector and the Frobenius norm of a matrix, respectively. $\hat{(\cdot)}$ is an estimate of $(\cdot)$, and $\text{diag}(\mathbf{x})$ is a diagonal matrix with $\mathbf{x}$ being its diagonal elements. $\mathbf{X}(:,i)$ and $\mathbf{X}(j,:)$ show the $i$-th column and the $j$-th row of $\mathbf{X}$, respectively. $\text{vec}(\mathbf{X})$ stands for the vector with the elements consists of columns of $\mathbf{X}$ stacked together.


\section{SIGNAL MODEL}
\label{sec:data model}
\begin{figure}[ht]
    \centering
    \includegraphics[width=0.8\linewidth]{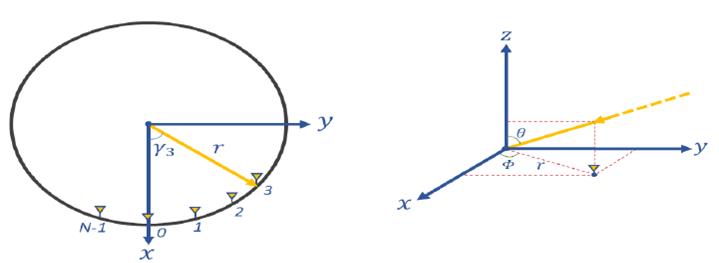}
    
    \caption{The considered UCA receiver model for DOA and MCM estimation.}
    \label{fig:fig1}
\end{figure}

Suppose that \( K \) narrowband far-field sources simultaneously impinge onto a uniform circular array (UCA) of omnidirectional sensors from angles of arrival \((\phi, \theta)\), where \( \phi \) denotes the azimuth angle, i.e., \( \phi \in [0, 360) \) degrees, measuring counterclockwise from the \( x \)-axis, and \( \theta \) denotes the elevation angle, i.e., \( \theta \in [0, 90] \) degrees, measuring down from the \( z \)-axis. The UCA geometry, with \( N \) equispaced sensors on the circumference of the array, is shown in Fig.~\ref{fig:fig1}  in the \( xy \)-plane. The marked arrow represents an example of a received signal geometrical description.

Therefore, by considering an array of $N$ sensors which are placed uniformly on a circle of radius $r$, the array output $\mathbf{x} \in \mathbb{C}^{N \times T}$ may be expressed as
\begin{equation}
\label{eq:eq1}
\mathbf{x}(t) = \mathbf{A}(\boldsymbol{\phi}, \boldsymbol{\theta}) + \mathbf{e}(t), \quad t = t_1, t_2, \ldots, t_T,
\end{equation}
where $\mathbf{A}(\boldsymbol{\phi}, \boldsymbol{\theta}) = [ \mathbf{a}(\phi_1, \theta_1), \mathbf{a}(\phi_2, \theta_2), \ldots, \mathbf{a}(\phi_K, \theta_K) ]$ is the array manifold/steering matrix, with
\[
\mathbf{a}(\phi_i, \theta_i) = \exp(j\xi_i \cos(\phi_i - \gamma_n)), \quad n = 0, 1, \ldots, N-1,
\]
denoting the steering vector of the source from $(\phi_i, \theta_i)$. Here, $\gamma_n = \frac{2\pi n}{N}$ is the displacement of the $n$th element of the array from the $x$ axis, $\xi_i = k_0 r \sin \theta_i$, with the wave number $k_0 = \frac{2\pi}{\lambda}$, whereas $\mathbf{s}(t) = [s_1(t), s_2(t), \ldots, s_K(t)]^T$ and $\mathbf{e}(t) = [e_1(t), e_2(t), \ldots, e_N(t)]^T$ denote the source data and an additive white Gaussian noise vector, respectively. Finally, $t$ indexes each snapshot and $T$ is the number of snapshots. The signal vector $\mathbf{s}(t)$ and the noise vector $\mathbf{e}(t)$ are assumed to be zero mean circularly symmetric and statistically independent processes.

\section{THE PROPOSED AUTOCALIBRATION METHOD}
\label{sec:proposed method}

By taking mutual coupling into consideration, the observation model in (\ref{eq:eq1}) can be reformulated as
\begin{equation}
\label{eq:eq2}
\mathbf{X} = \mathbf{C} \mathbf{A}(\phi, \theta) \mathbf{S} + \mathbf{E},
\end{equation}
where \(\mathbf{X} = [\mathbf{x}(t_1), \mathbf{x}(t_2), \ldots, \mathbf{x}(t_T)]\), \(\mathbf{S} = [\mathbf{s}(t_1), \mathbf{s}(t_2), \ldots, \mathbf{s}(t_T)]\), and \(\mathbf{E} = [\mathbf{e}(t_1), \mathbf{e}(t_2), \ldots, \mathbf{e}(t_T)]\) are the array output, as well as the source and noise data matrices, respectively. Furthermore, \(\mathbf{C}\) denotes the mutual coupling matrix which is assumed to have a complex symmetric circular Toeplitz structure such that  
\begin{equation}
\label{eq:eq3}
\mathbf{C} = \text{toeplitz}(\tilde{\mathbf{c}}, \tilde{\mathbf{c}}) \in \mathbb{C}^{N,N},
\end{equation}
where
\[
\mathbf{\tilde{c}} = [c_1, c_2, \ldots, c_L, c_{L-1}, \ldots, c_3, c_2] \in \mathbb{C}^{1,N}, \quad L = \frac{N}{2} + 1, \text{ for } N \text{ even,}
\]
and
\[
\tilde{\mathbf{c}} = [c_1, c_2, \ldots, c_L, \ldots, c_3, c_2] \in \mathbb{C}^{1,N}, \quad L = \left(\frac{N + 1}{2}\right), \text{ for } N \text{ odd.}
\]
Here, $L$ denotes the degrees of freedom of the MCM, i.e., the number of the unknown elements in the first row of the matrix $\mathbf{C}$. We note that because of the inverse relationship of mutual coupling and distance between each pairs of sensors \( 0 < |c_L| < \ldots < |c_2| < |c_1| = 1 \).

Algorithm 1 summarizes the proposed method for estimating the DOA and the MCM. Each substep will be elaborated upon in the following subsections.

\begin{algorithm}
\caption{The proposed SVD-based sparse WB method}
\begin{algorithmic}[1]
    \State Assign an initial MCM such that $\mathbf{C}_0 = \mathbf{I}_{N \times N}$.
    \State Choose the number of zooming steps and number of bands to use for both azimuth and elevation; select a suitable $\alpha$ for Eq. (\ref{eq:eq11}).
    \State Compute the number of sources and take the SVD of Eq. (\ref{eq:eq2}).
    \State Compose the DOA integrated wideband dictionary with elements defined in Eq. (\ref{eq:eq6}).
    \State Estimate the azimuth and elevation (as stated in Subsection \ref{subsec:proposed 2}) using Eq. (\ref{eq:eq10}).
    \State Estimate the MCM as detailed in Subsection \ref{subsec:proposed 3}.
    \State Form the DOA estimates on the active bands (composing of the zoomed dictionary given in Eq. (\ref{eq:eq6})), and then using Eq. (\ref{eq:eq10}) include the MCM in step 6. The iterative zooming procedure enables discarding the non-activated parts of each dictionary.
    \State Estimate the MCM as stated in Subsection \ref{subsec:proposed 3}\\
    Repeat steps 7 and 8 until obtaining the desired resolution.
\end{algorithmic}
\end{algorithm}

\subsection{Sparse SVD-based Representation}
\label{subsec:proposed 1}

The overall complexity of the proposed method may be notably reduced by projecting the measured data onto the \(K\)-dimensional signal subspace. This can be formed by the \(K\) dominant singular vectors of \(\mathbf{X}\). Form the singular value decomposition (SVD) as:
\begin{equation}
    \mathbf{X} = \mathbf{U}_s \Sigma_s \mathbf{V}_s^H + \mathbf{U}_c \Sigma_c \mathbf{V}_c^H,
\end{equation}
where \(\mathbf{U}_s \in \mathbb{C}^{N \times K}\) and \(\mathbf{V}_s \in \mathbb{C}^{T \times K}\) are the singular vectors corresponding to the \(K\) largest singular values, and \(K\) represents the number of signals (sources). To do this, the number of sources, \(K\), must also be estimated. Here, this is done by forming the resulting solution over a set of potential \(K \in \{1, 2, \ldots\}\) and then selecting the estimated \(K\) as the one resulting in the best signal modeling when combined with a BIC penalty to compensate for the measuring model order \cite{ref25}. Let $\mathbf{X}_{\text{SV}} \triangleq \mathbf{X}V_{s}$, $\mathbf{S}_{\text{SV}} \triangleq \mathbf{S}V_{s}$, and $\mathbf{E}_{\text{SV}} \triangleq \mathbf{E}V_{s}$, such that
\begin{equation}
\label{eq:eq5}
\mathbf{X}_{\text{SV}} = \mathbf{C}\mathbf{A}(\boldsymbol{\phi},\boldsymbol{\theta})\mathbf{S}_{\text{SV}} + \mathbf{E}_{\text{SV}},
\end{equation}

Indeed, by using the \(N \times K\) dimensional matrix \(\mathbf{X}_{\text{SV}}\) in place of \(\mathbf{X}\), one may significantly reduce the resulting computational complexity. The reduction will be significant in case of a small number of sources and/or a large number of time samples.

\subsection{DOA Estimation}
\label{subsec:proposed 2}

\begin{figure}[ht]
    \centering
    \includegraphics[width=0.5\linewidth]{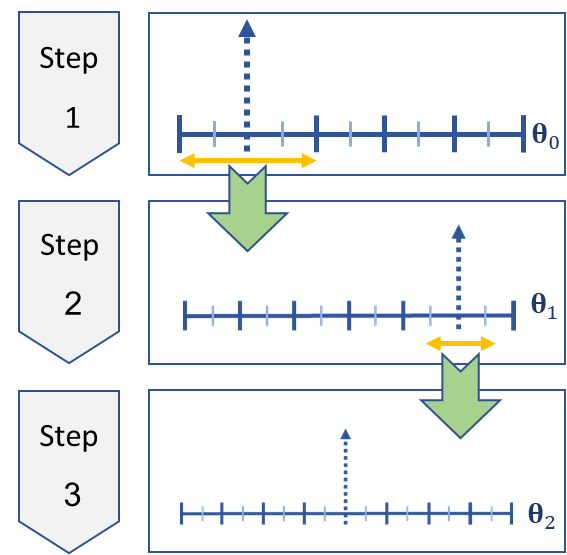}
    \caption{The process of zooming on intervals/bands of angles.}
    \label{fig:zooming_steps}
\end{figure}

Proceeding to determine the DOAs, let $(\phi_i,\theta_i)_{i=1}^{Q}$ for $i = 1,2,...,Q$, stands for the set of potential DOA candidates, with $Q$ indicating the number of candidates. Denote the dictionary over the set of candidates $\mathbf{A}(\hat{\phi},\hat{\theta}) = [\mathbf{a}(\hat{\phi}_1,\hat{\theta}_1), \mathbf{a}(\hat{\phi}_2,\hat{\theta}_2),...,\mathbf{a}(\hat{\phi}_Q,\hat{\theta}_Q)]$, where $\mathbf{a}(\hat{\phi}_i,\hat{\theta}_i) = [\mathbf{a}_0(\hat{\phi}_i,\hat{\theta}_i),\mathbf{a}_1(\hat{\phi}_i,\hat{\theta}_i),...,\mathbf{a}_{N-1}(\hat{\phi}_i,\hat{\theta}_i)]^T$ and each candidate is formed using the integrated wideband element, $\mathbf{a}_{n_r}(\phi,\theta)$, such that
\begin{equation}
\label{eq:eq6}
\mathbf{a}_{n_r}(\phi_i,\theta_i) = \int_{\phi_i}^{\phi_{i+1}} \int_{\theta_i}^{\theta_{i+1}} \exp\left(jk_0r\sin \theta \cos\left(\phi - \frac{2\pi n_r}{N}\right)\right) d\theta d\phi,
\end{equation}
with $n_r = 0,1,...,N - 1$, and where $[\phi_i,\phi_{i+1})$ and $[\theta_i,\theta_{i+1})$ are the intervals for the $i$th azimuth and elevation candidate. Indeed, instead of using a finely spaced narrowband dictionary of the candidate DOAs, the 2-D angles domain, i.e., azimuth $\phi : [0^\circ, 360^\circ)$ and elevation $\theta : [0^\circ, 90^\circ)$, is here divided into $B_1$ and $B_2$ continuous bands, respectively. As long as $B_1$ and $B_2$ are selected to be sufficiently large, this process would reduce the risk of any off-grid components and provide an initial coarse estimation of the angle of arrivals. Since all the bands are covered in the dictionary, no power is off-grid, which in turn avoids a non-sparse solution due to dictionary mismatch. In its initial step, this will yield a coarse estimation of the regions of interest, whereafter one may discard the non-activated bands and form the refined dictionary over the activated bands. This process may then be repeated until one reaches the desired resolution. In other words, instead of making a dictionary over a fine grid to limit the performance degradation due to off-grid effects, thereby necessitating a substantial computational burden, we use the integrated dictionary to determine and refine the estimate of the regions where sources are present.
The process of zooming in some steps (here, 3 steps) is illustrated in Fig.~\ref{fig:zooming_steps}, for the case of a single angle \(\theta\). As shown in \cite{ref20}, the average computation time of employing the wideband dictionary is substantially less than what would be required by the corresponding narrowband dictionary. The main difference of this zooming with respect to the earlier zooming methods is in utilizing an integrated dictionary for bands which significantly improves the performance.
Suppose the arrow in Fig.~\ref{fig:zooming_steps} shows the location of one of the sources. Then, one may describe the process of zooming in three stages as follows:
\begin{enumerate}
    \item Initially, one forms an integrated dictionary over the coarse grid (\(\mathbf{\theta}_0\)).
    \item Then, one identifies the regions that may contain a source (as an example, the right side of \(\mathbf{\theta}_1\) is a possible region).
    \item Finally, one composes an integrated dictionary with finer grids over the possible regions identified in step 2 (the third step is shown in the \(\mathbf{\theta}_2\) region).
\end{enumerate}

\textbf{Remark:} In Appendix A, we introduce an approach to form the 2-D integral in equation (6).

Proceeding, the azimuth and elevation estimation are formulated using the LASSO \cite{ref26}, which solves the resulting penalized regression problem. To this end, let \(\mathbf{\hat{A}} = \mathbf{\hat{C}}\mathbf{A}(\hat{\phi},\hat{\theta}) \in \mathbb{C}^{M \times Q}\) and then
\begin{equation}
\label{eq:eq7}
\mathbf{D} = 
\begin{bmatrix}
\mathbf{\hat{A}} \\
\vdots \\
\mathbf{\hat{A}}
\end{bmatrix}_{MK \times Q}
\end{equation}

Then,
\begin{equation}
\label{eq:eq8}
\begin{bmatrix}
\mathbf{X}_{\text{SV}}(:,1) \\
\vdots \\
\mathbf{X}_{\text{SV}}(:,K)
\end{bmatrix}
=
\mathbf{D}\mathbf{\hat{s}} +
\begin{bmatrix}
\mathbf{N}_{\text{SV}}(:,1) \\
\vdots \\
\mathbf{N}_{\text{SV}}(:,K)
\end{bmatrix}
\end{equation}

Let \(\mathbf{x} = \text{vec}(\mathbf{X}_{\text{sv}}) \in \mathbb{C}^{MK}\) and \(\mathbf{n} = \text{vec}(\mathbf{N}_{\text{sv}}) \in \mathbb{C}^{MK}\). Then,

\begin{equation}
\label{eq:eq9}
\mathbf{x} = \mathbf{D}_{MK \times Q}\mathbf{\hat{s}} + \mathbf{n},
\end{equation}

allowing the LASSO formulation\textsuperscript{1}, i.e., \(\mathbf{\hat{s}}\), results in

\begin{equation}
\label{eq:eq10}
\mathbf{\hat{s}} = \arg\min_{\mathbf{\hat{s}}} \left\{ \left\| \mathbf{x} - \mathbf{D}\mathbf{\hat{s}} \right\|_{2}^{2} + \gamma \left\| \mathbf{\hat{s}} \right\|_{1} \right\},
\end{equation}

where \(\mathbf{\hat{s}} \in \mathbb{C}^{Q \times 1}\) and \(\gamma\) is a user chosen parameter controlling the sparsity of the solution, here being selected as

\begin{equation}
\label{eq:eq11}
    \gamma = \alpha \max_i |\mathbf{a}_i^H \mathbf{x}|,
\end{equation}
where \(\mathbf{a}_i\), \(i = 1, 2, \ldots, Q\) are columns of the \(\mathbf{D}\). The constant \(\alpha\) is a user-chosen parameter allowing the performance of each dictionary to be evaluated (as in the simulation results) by varying values of this parameter within the range \(0 < \alpha \leq 1\), with \(\gamma_{max} = \alpha \max_i |\mathbf{a}_i^H \mathbf{x}|\) denoting the smallest tuning parameter value that gives a solution with the coefficients of zero \cite{ref20, ref26}.

\subsection{MCM Estimation}
\label{subsec:proposed 3}
Proceeding, we estimate the mutual coupling matrix coefficients by utilizing the complex symmetric circular Toeplitz structure of the MCM in a UCA. Suppose \(\mathbf{R}_x\) denotes the array output covariance matrix and let \(\lambda_n\) and \(\mathbf{e}_n\) denote its eigenvalues and the corresponding eigenvectors, respectively. Then,
\begin{equation}
    \mathbf{R}_x = \sum_{n=1}^N \lambda_n \mathbf{e}_n \mathbf{e}_n^H = \mathbf{E}_s \Lambda_s \mathbf{E}_s^H + \mathbf{E}_n \Lambda_n \mathbf{E}_n^H,
\end{equation}
where \(
\mathbf{E}_s = [\mathbf{e}_1, \mathbf{e}_2, \ldots, \mathbf{e}_K] \in \mathbb{C}^{N \times K}
\)
denotes the signal subspace containing the principal eigenvectors corresponding to the \(K\) maximum eigenvalues, with the noise subspace \(\mathbf{E}_n\) contains the remaining \(N-K\) eigenvectors.

The signal subspace spans the same space as the array manifold matrix, i.e.,
\(\text{span}\{\mathbf{E}_s\} \perp \text{span}\{\mathbf{CA(\phi, \theta)}\}\)
. Moreover, the signal subspace and noise subspace are orthogonal such that \(\text{span}\{\mathbf{E}_s\} \perp \text{span}\{\mathbf{E}_n\}\), implying that 
\(\text{span}\{\mathbf{E}_s\} \perp \text{span}\{\mathbf{E}_n\}\), Implying that 
\(\text{span}\{\mathbf{E}_n\} \perp \text{span}\{\mathbf{CA(\phi, \theta)}\}\)
. \\
By supposing that we have DOAs (as obtained in Subsection \ref{subsec:proposed 2}), the MCM could be estimated as the matrix \(\mathbf{C}\) minimizing:

\begin{equation}
\label{eq:eq13}
J = \left\| \mathbf{E}_n^H \mathbf{C} \mathbf{A}(\phi, \theta) \right\|_F^2 = \sum_{k=1}^K \left\| \mathbf{E}_n^H \mathbf{C} \mathbf{a}(\phi_k, \theta_k) \right\|^2
\end{equation}

where \(\| \cdot \|_F\) denotes the Frobenius norm.

By using the circular symmetry of the mutual coupling matrix, one may instead estimate the vector which constructs the mutual coupling matrix, i.e., by rewriting \(
\mathbf{C} \mathbf{a}(\phi, \theta) = \mathbf{F}\{\mathbf{a}(\phi, \theta)\} \mathbf{c} \in \mathbb{C}^{L \times 1}\)
 where \(\mathbf{F}\) denotes the resulting transform matrix, and \(\mathbf{c} = [c_1, c_2, \ldots, c_L]^T \in \mathbb{C}^L\). By extending the relation of the mutual coupling introduced in \cite{ref23,ref24} to 2-D and exploiting the symmetric circulant property of \(\mathbf{C}\), the matrix \(\mathbf{F}\{\mathbf{a}(\phi, \theta)\}\) may be defined as:
\begin{equation}
\mathbf{F}\{\mathbf{a}(\phi, \theta)\} = \mathbf{F}_1\{\mathbf{a}(\phi, \theta)\} + \mathbf{F}_2\{\mathbf{a}(\phi, \theta)\} + \mathbf{F}_3\{\mathbf{a}(\phi, \theta)\} + \mathbf{F}_4\{\mathbf{a}(\phi, \theta)\}
\end{equation}

\begin{equation}
\begin{aligned}
\left[\mathbf{F}_1\{\mathbf{a}(\phi,\theta)\}\right]_{i,j} &= 
\begin{cases}
\left[\mathbf{a}(\phi,\theta)\right]_{i+j-1,1} & \text{if } i+j \leq N + 1, \\
0 & \text{otherwise},
\end{cases} \\
\left[\mathbf{F}_2\{\mathbf{a}(\phi,\theta)\}\right]_{i,j} &= 
\begin{cases}
\left[\mathbf{a}(\phi,\theta)\right]_{j-i+1,1} & \text{if } i \geq 2, \\
0 & \text{otherwise},
\end{cases} \\
\left[\mathbf{F}_3\{\mathbf{a}(\phi,\theta)\}\right]_{i,j} &= 
\begin{cases}
\left[\mathbf{a}(\phi,\theta)\right]_{N+i+j-1,1} & \text{if } i < j \leq p, \\
0 & \text{otherwise},
\end{cases} \\
\left[\mathbf{F}_4\{\mathbf{a}(\phi,\theta)\}\right]_{i,j} &= 
\begin{cases}
\left[\mathbf{a}(\phi,\theta)\right]_{i+j-N-1,1} & \text{if } 2 \leq j \leq p, i+j \geq N + 2, \\
0 & \text{otherwise}.
\end{cases}
\end{aligned}
\end{equation}

where \( i = 1, 2, \ldots, N \), \( j = 1, 2, \ldots, L \), and \( p = \left\lfloor \frac{N + 1}{2} \right\rfloor \). Applying this transformation to (\ref{eq:eq13}) yields

\begin{equation}
\label{eq:eq16}
J = \sum_{k=1}^K \mathbf{c}^H \mathbf{F}^H\{\mathbf{a}(\phi_k,\theta_k)\} \mathbf{E}_n \mathbf{E}_n^H \mathbf{a}(\phi_k,\theta_k)\mathbf{F}\{a(\phi_k,\theta_k)\} \mathbf{c} = \mathbf{c}^H \mathbf{U}(\phi,\theta) \mathbf{c},
\end{equation}

where

\begin{equation}
\mathbf{U}(\phi,\theta) = \sum_{k=1}^K  \mathbf{F}^H\{\mathbf{a}(\phi_k,\theta_k)\} \mathbf{E}_n\mathbf{E}_n^H \mathbf{F}\{\mathbf{a}(\phi_k,\theta_k)\}^H \in \mathbb{C}^{L \times L},
\end{equation}

By imposing the constraint that \( c_1 = 1 \), (\ref{eq:eq16}) may be used as the cost function to be minimized in order to obtain the desired mutual coupling vector, i.e.,

\begin{equation}
\mathbf{c} = \mathbf{e}_{\min}\left[\mathbf{U}(\phi,\theta)\right],
\end{equation}

where \( \mathbf{e}_{\min}[\cdot] \) denotes the eigenvector corresponding to the smallest eigenvalue.

\section{Simulation Results}
\label{sec:simulations}
In this section, some simulations will be carried out to investigate the performance of the proposed method. We consider a UCA with \( N = 15 \) sensors, the radius of the UCA being \( r = \lambda \), where \( \lambda \) denotes the center wavelength of the narrow-band signals. We further assume three far-field sources from angle of arrivals \( (18.3^\circ, 243.4^\circ) \), \( (83.6^\circ, 60^\circ) \), and \( (73.9^\circ, 357.8^\circ) \) for which \( T = 200 \) snapshots are measured. The signals and the additive sensor noises are stationary, mutually uncorrelated, and zero mean. The signals are modeled as circularly symmetric Gaussian processes with variance \( \sigma_i^2 \) for the \( i \)-th process. The corrupting noise is assumed to be i.i.d. white circularly symmetric Gaussian process with variance \( \sigma^2 \), such that the input SNR of the \( i \)-th signal is \( 10 \log(\sigma_i^2/\sigma^2) \). The coupling coefficient is set to be \( c_2 = 0.79 + j0.432 \) and \( c_3 = 0.35 + j0.16 \).

Fig.~3 presents the probability of correctly estimating the number of angle of arrivals using an integrated wideband dictionary. Suppose \( B^{\theta}_i \) and \( B^{\phi}_i \) denote the number of bands over elevation and azimuth, respectively. Also, we have \( i = 1, 2, 3 \) for each stage of the dictionary. We here consider for the zooming steps of the elevation \( B^{\theta}_1 = 30 \), \( B^{\theta}_2 = 10 \), \( B^{\theta}_3 = 3 \), and for the azimuth \( B^{\phi}_1 = 120 \), \( B^{\phi}_2 = 10 \), and \( B^{\phi}_3 = 3 \) (the grey curve), and \( B^{\theta}_1 = 30 \), \( B^{\theta}_2 = 5 \), \( B^{\theta}_3 = 3 \), \( B^{\phi}_1 = 120 \), \( B^{\phi}_2 = 5 \), and \( B^{\phi}_3 = 3 \) (the brown curve). In this figure, the correct model order is given for varying values of \( \alpha \) for (\ref{eq:eq11}). As is clear from Fig.~3, using \( \alpha \leq 0.4 \) for the three-stage dictionary (the grey curve) is a suitable region for utilizing the proposed method.

\begin{figure}[ht]
    \centering
    \includegraphics[width=0.8\linewidth]{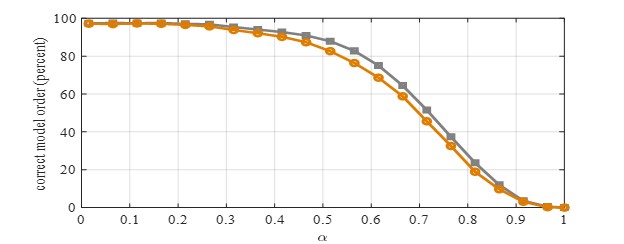}
    \caption{Choosing suitable \( \alpha \) for equation (\ref{eq:eq11}).}
    \label{fig:suitable_alpha}
\end{figure}

To evaluate the precision of DOA estimation methods, the average root-mean-square errors (RMSEs) of the elevation and the azimuth estimation are used as follows:
\begin{equation}
\text{RMSE} = \sqrt{\frac{1}{KM} \sum_{k=1}^K \sum_{m=1}^M \left[ (\theta_k^m - \theta_k)^2 + (\phi_k^m - \phi_k)^2 \right]},
\end{equation}
for \( M = 500 \) number of Monte Carlo simulations, where \( (\theta_k^m, \phi_k^m) \) denotes the estimate of \( (\theta_k, \phi_k) \) in the \( m \)-th Monte Carlo run (out of \( M \) runs). Fig. 4 shows the resulting RMSE when the SNR varies from 0 dB to 20 dB. The results are compared with the LASSO estimates resulting when the dictionary is composed of discrete prolate spheroidal sequences (DPSS) \cite{ref27,ref28}, and to the 2-D extension of the method in \cite{20}, here termed the Iterative-MUSIC estimator. This Iterative-MUSIC is found to have the highest computational complexity while offering the worst performance of the
discussed methods.  

\begin{figure}[ht]
    \centering
    \includegraphics[width=0.8\linewidth]{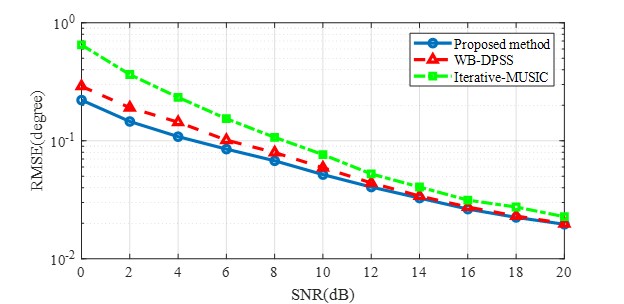}
    \caption{The RMSE of angles of estimation as a function of the SNR.}
    \label{fig:fig4}
\end{figure}
\begin{figure}[ht]
    \centering
    \includegraphics[width=0.8\linewidth]{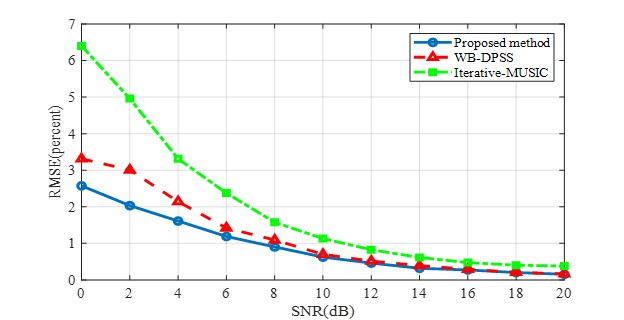}
    \caption{The RMSE of the mutual coupling coefficients estimation as a function of the SNR.}
    \label{fig:fig5}
\end{figure}

Fig.~\ref{fig:fig5} shows the RMSE of coupling coefficients, being defined as

\begin{equation}
    \text{RMSE}_{\text{coupling coeff.}} = \frac{\sqrt{\sum_{m=1}^M \| \mathbf{\hat{c}}^m - \mathbf{c} \|^2}}{\| \mathbf{c} \|} \times 100\%,
\end{equation}
where \( \mathbf{\hat{c}}^m \) is the estimation of coupling coefficient vector, \( \mathbf{c} \), in the \( m \)-th Monte Carlo experiment. As shown in Fig.~\ref{fig:fig5}, the coupling coefficients estimation of the proposed method is very effective and also has a very high precision.
\begin{figure}[ht]
    \centering
    \includegraphics[width=0.8\linewidth]{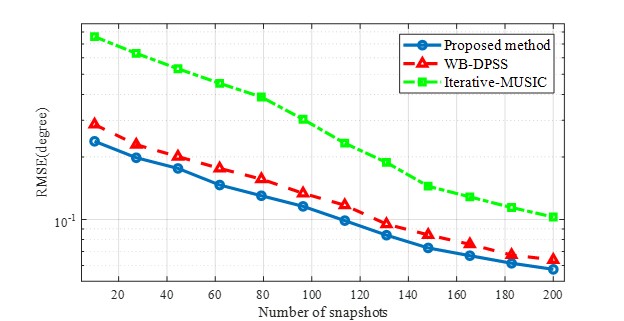}
    \caption{The RMSE of the angle estimates as a function of the number of the measured snapshots.}
    \label{fig:fig6}
\end{figure}
In order to compare the performance in terms of the number of snapshots, we fix the SNR at \( 5 \) dB, \( M = 200 \) and vary the snapshot number. The results are shown in Fig.~\ref{fig:fig6}. The other parameters are set as before. As seen in Fig.~\ref{fig:fig3} to Fig.~\ref{fig:fig6}, the proposed method offers preferable performance of the three methods.

\section{Conclusion}
\label{sec:conclusion}
In this work, we have introduced a self-calibration algorithm which is able to jointly estimate the 2-D DOAs and the MCM for a UCA configuration. The DOA estimation step is formed using a sparse representation framework in combination with an integrated dictionary elements, which span bands of the desired parameter space. The MCM estimation step exploits the circular symmetry of the mutual coupling matrix resulting for UCAs. This method has the advantages of allowing for an initial coarse gridding of the dictionary, without the risk of suffering from off-grid effects, as well as not posing any constraints on the coupling matrix, nor resulting in any ambiguity in the resulting DOA estimates. Several simulation results confirm the superior performance of the proposed method.


%

\appendices
\section{Simplifying the 2-D integral of the dictionary elements in order to allow for non-numerical
computations}
To simplify the 2-D integration in (\ref{eq:eq6}) and attain a formula for computing the integration, one may use the Taylor series expansion for \( e^z \) about \( z = 0 \), i.e.,
\begin{equation}
e^z = \sum_{n=0}^{\infty} \frac{z^n}{n!}
\end{equation}

\begin{align}
\nu \triangleq & \int_{\phi_i}^{\phi_{i+1}} \int_{\theta_i}^{\theta_{i+1}} e^{j\alpha r \sin\theta \cos(\phi - \lambda)} d\theta d\phi \notag \\
= & \int_{\phi_i}^{\phi_{i+1}} \int_{\theta_i}^{\theta_{i+1}} \sum_{n=0}^{\infty} \frac{\left(\alpha \cos(\phi - \lambda)\right)^n}{n!} \sin^n \theta d\theta d\phi \notag \\
= & \int_{\phi_i}^{\phi_{i+1}} \sum_{n=0}^{\infty} \left(\frac{\alpha^n}{n!} \cos^n(\phi - \lambda) \int_{\theta_i}^{\theta_{i+1}} \sin^n \theta d\theta\right) d\phi \notag \\
= & \sum_{n=0}^{\infty} \frac{\alpha^n}{n!} \int_{\phi_i}^{\phi_{i+1}} \cos^n(\phi - \lambda) d\phi \int_{\theta_i}^{\theta_{i+1}} \sin^n \theta d\theta \notag \\
= & \sum_{n=0}^{\infty} \frac{\alpha^n}{n!} \int_{-\lambda_{n}}^{\phi_{i+1}-\lambda_{n}} \cos^n \phi d\phi \int_{\theta_i}^{\theta_{i+1}} \sin^n \theta d\theta \label{eq:myeq}
\end{align}

Computing the first and the second terms yields

\scalebox{0.8}{
\begin{minipage}{1.0\linewidth}
\begin{equation}
\nu = \left( \int_{\phi_i}^{\phi_{i+1}-\lambda_n} \phi d\phi \right) \left( \int_{\theta_i}^{\theta_{i+1}} d\theta \right) + \alpha \left( \sin \int_{\phi_i}^{\phi_{i+1}-\lambda_n} d\phi \right) \left( -\cos \int_{\theta_i}^{\theta_{i+1}} \theta d\theta \right) + \sum_{n=2}^{\infty} \frac{\alpha^n}{n!} \int_{\phi_i}^{\phi_{i+1}-\lambda_n} \cos^n \phi d\phi \int_{\theta_i}^{\theta_{i+1}} \sin^n \theta d\theta
\end{equation}
\end{minipage}
}

Integration by parts for the third term of the above equation yields
\begin{equation}
\int \sin^n \theta d\theta = -\frac{1}{n} \sin^{n-1} \theta \cos \theta + \frac{n-1}{n} \int \sin^{n-2} \theta d\theta
\end{equation}
and
\begin{equation}
\int \cos^n \phi d\phi = \frac{1}{n} \cos^{n-1} \phi \sin \phi + \frac{n-1}{n} \int \cos^{n-2} \phi d\phi
\end{equation}

One may thus avoid a numerical integration by computing the recursive factors in the terms. To do this, the number of terms (for this work, computing about $\sim 10$ terms suffices and gives the same precision as the built-in \texttt{integral2} function in Matlab) in this series could be computed to get the required precision.

\section*{Acknowledgment}

The first author would like to thank Dr. Andreas Jakobsson (Dept. of Mathematical Statistics, Lund University) because of his insightful comments and discussions.

\ifCLASSOPTIONcaptionsoff
  \newpage
\fi



%

\bibliographystyle{IEEEtran}
\bibliography{references}




\end{document}